%% file: ett1.tex
\documentclass[14]{article}

\usepackage{bussproofs}
\usepackage{mathrsfs}
\usepackage{amsmath}
\usepackage{amsfonts}
\usepackage{stmaryrd}
\usepackage{MnSymbol}
\usepackage{amsthm}

\DeclareMathAlphabet{\mathpzc}{OT1}{pzc}{m}{it}
\newtheorem{prop}{Proposition}
\newtheorem{cor}[prop]{Corollary}
\newtheorem{lemma}[prop]{Lemma}
\newtheorem{thm}[prop]{Theorem}
\theoremstyle{definition}
\newtheorem{dfn}[prop]{Definition}

\EnableBpAbbreviations

\title{On Extensionality of $\lstar$}
\author{Andrew Polonsky}

\usepackage{natbib}

\input{amslambda}

\renewcommand{\cite}{\citet}
\begin{document}

\maketitle

\begin{abstract}
We prove an extensionality theorem for the ``type-in-type''
dependent type theory with $\Sigma$-types.  We suggest that the
extensional equality type be identified with the logical equivalence
relation on the free term model of type theory.
\end{abstract}

\section{Introduction}
The \emph{Extensionality Theorem} of simple type theory
states that all definable terms preserve a given relation $R_A$ on the
free model of type theory, provided this relation is logical.
In this context, \emph{logical} means that the relation is
generated in a very specific way from the base types by
induction on the type structure.  In particular, for the function
type $A \to B$, the relation is to be given by
\[ R_{A \to B} f f' \;\iff\; (\forall a a' : A) R_A a a' \to R_B (f a)
(f' a') \]

The theorem has its origins in the Tarski--Sher thesis on the
invariance of truth-valued operations (\cite{Feferman}, \cite{Tarski},
\cite{Sher}).   The first application of what came to be known as
\emph{logical relations} technique was given by \cite{Gandy}
in the proof of relative consistency of the axiom of extensionality
in Church's theory of types.  However, it is only fairly recently that
William \cite{Tait} suggested that the notion of extensional equality
in type theory be identified with the meta-level equivalence relation
between terms defined by induction on type structure.  By a logical
relations argument, Tait proceeded to show that every term is indeed
extensional, ie, preserves the semantic equivalence relation.  As a
consequence, every \emph{closed} term is related to iself by this relation, giving a
a computational justification for introducing the reflexivity operator
for extensional equality.

Tait's ideas remain relatively little-known.  In light of the recent
attention received by issues of extensional identity, we think this is a good
time to explicitly announce the following
\begin{description}
\item[Extensionality Thesis.] The extensional equality of type theory
  is the logical equivalence relation between elements of the term
  model defined by induction on type structure.  
\end{description}

In this note, we shall generalize Tait's extensionality theorem to dependent
type theory with the universe of all types being itself a type.  This
system is known as $\lstar$.  It is a pure type system which includes
$\Sigma$-types in addition to $\Pi$.
(A \emph{$\Pi\Sigma$-system}, in the lexicon of \cite{Terlouw}.)

Our result is closely related to the well-known theory of
\emph{parametricity}, which generalizes Reynolds'
Abstraction Theorem for the polymorphic lambda calculus to the
dependent case.  For the PTS formulation of dependent type theory,
the general result is proved by \cite{Bernardy}.  For the Logical
Frameworks formulation, it is proved by \cite{Sojakova}.

The central difference between the above results and ours consists in
the treatment of universes.  In parametricity theory, one associates
to every type a relation $R_A : A \to A \to \btype$ (which in general can
have arbitrary arity).  In the case of the universe $*$, a relation
between two types $A,B : *$ is just that --- a term of type $A \to B
\to \btype$.  So $R_* A B$ is just the type of relations between $A$
and $B$.

In our case, we want the relation on the universe to be a (1-dimensional)
\emph{equivalence} of types, which means that a term of type
$R_* A B$ is not merely a relation, but a relation satisfying certain additional
properties.  We give a formulation of these properties in the
language of Induction--Recursion (\cite{Dybjer}), and prove
a strict extensionality theorem for $\lambda *$.

\subsection{The simply typed case} \label{s:stt}

We begin by recalling an elementary fact about the simply typed $\lambda$-calculus.

The syntax of simple types and typed terms is as follows:
\begin{align*}
\types &= o \mid \types \to \types \mid \types \times \types\\
\terms &= x \mid \lambda x{:}\types. \terms \mid \terms \terms\\
&\phantom{= x} \; \mid (\terms,\terms) \mid \pi_1 \terms \mid \pi_2 \terms
\end{align*}

A model of $\lsimple$ consists of a family of sets
$\setof{X_A \mid A \in \types}$ where
\begin{align*}
  X_{A \to B} &\sse X_A^{X_B}\\
  X_{A \times B} &\sse X_A \times X_B
\end{align*}
are such that $X_{A \to B}$ is closed under abstraction of terms of
type $B$ over variables of type $A$, and $X_{A \times B}$ is closed under pairs of definable elements of $X_A$ and $X_B$.

The interpretation of types is given by
\[ \semof{A} = X_A \]

The interpretation of terms is parametrized by an environment
$\rho = \setof{\rho_A : V_A \to  X_A}$, assigning elements of the domain to the
free variables of the term.

Let $\env$ be the set of such collections of functions.

A term $t:A$ is interpreted as a map $\semof{t} : \env \to \semof{A}$.
We write $\semof{t}_\rho$ for $\semof{t}(\rho)$.
The definition of $\semof{t}_\rho$ is given by induction:
\begin{align*}
  \semof{x:A}_\rho &= \rho_A(x)\\
  \semof{s t}_\rho &= \semof{s}_\rho \semof{t}_\rho\\
  \semof{\lambda x{:}A.t}_\rho &= (a \mapsto \semof{t}_{\rho,x := a})\\
  \semof{(s,t)}_\rho &= (\semof{s}_\rho,\semof{t}_\rho)\\
  \semof{\pi_i t}_\rho &= a_i, \text{ where } \semof{t}_\rho=(a_1,a_2) \in
\semof{A_1 \times A_2}
\end{align*}

A relation $R = \setof{R_A : \semof{A} \to \semof{A} \to \btype \mid A \in \types}$ is said to be \emph{logical} if
\begin{align*}
  R_{A \to B} f f' &\iff \forall a a'{:}X_A. R_A a a' \RA R_B (f a) (f' a')\\
  R_{A \times B} (a,b) (a',b') &\iff R_A a a' \land R_B b b'
\end{align*}

\begin{thm}
(Extensionality Theorem)  Let $R$ be logical.  Suppose that
$t$ is a typed term:
\[x_1 : A_1, \dots, x_n : A_n \vdash t : T\]
and let there be given
\[a_1, a_1' \in \semof{A_1}, \dots, a_n, a_n' \in \semof{A_n}\]
Then
\[a_1 R_{A_1} a_1', \dots, a_n R_{A_n} a_n' \then \semof{t}_{\vec x := \vec a}
R_T\semof{t}_{\vec x := \vec a'}\]
\end{thm}

In other words, every typed $\lambda$-term induces a function which maps related elements to related elements.  As a corallary, we get that a closed term
$t \in \Lambda^0(A)$ is $R_A$-related to itself.

We also note that if a given relation $R_o$ on the basic type is reflexive (symmetric, transitive), then its logical extension to the full type structure is also reflexive (symmetric, transitive).  In particular, any equivalence given on $X_o$ can be extended to the interpretation of all types $X_A$ by the logical conditions.  Then the elements in the model which are defined by $\lambda$-terms will preserve the equivalence relation on the corresponding types.

The proof of the above theorem proceeds by induction on the structure of derivation that $t : T$.  We do abstraction case as an example.  If $t = \lambda x{:}A.t'$, we have\comment{(by inversion)}
\begin{prooftree}
  \AXC{$x_1:A_1,\dots,x_n:A_n,x:A \vdash t' : B$}
  \RL{Abs}
  \UIC{$x_1:A_1,\dots,x_n:A_n \vdash \lambda x{:}A. t' : A \to B$}
\end{prooftree}

Let $(a_1,\dots,a_n), (a_1',\dots,a_n') : A_1 \times \cdots \times A_n$
be such that $a_i R_{A_i} a_i'$.
Assume $a,a' : A$ are given, and suppose that $a R_A a'$.

By induction hypothesis,
\[\semof{t'}_{x_1\dots x_n x:=a_1\dots a_n a} R_B \semof{t'}_{x_1\dots x_n x:=a_1'\dots a_n' a'}\]
which can be rewritten as
\begin{align*}
  \semof{\lambda x.t'}_{x_1\dots x_n:=a_1\dots a_n}(a) R_B \semof{\lambda x.t'}_{x_1\dots x_n:=a_1'\dots a_n'}(a')
\end{align*}
Since $a,a'$ were arbitrary, and $R_{A\to B}$ is logical, it follows that
\[R_{A\to B} \semof{\lambda x.t'}_{x_1\dots x_n:=a_1\dots a_n}
          \semof{\lambda x.t'}_{x_1\dots x_n:=a_1'\dots a_n'}\]
The other cases are treated similarly.

We note that the structure of the proof that $t R_T t$ recapitulates rather precisely the structure of $t$ itself.  In particular, the theorem is completely constructive.  Anticipating the dependent development below, consider a constructive reading of the theorem's statement:
\begin{align*}
&\mbox{From a\; proof $a_1^*$ that $a_1 R_{A_1} a_1'$}\\
&\quad \mbox{and a proof $a_2^*$ that $a_2 R_{A_2} a_2'$}\\
&\quad \cdots\\
&\quad \mbox{and a proof $a_n^*$ that $a_n R_{A_n} a_n'$}\\
&\mbox{Get a proof $t(a_1^*,\dots,a_n^*)$}\\
&\quad \mbox{that $t(a_1,\dots,a_n) R_T t(a_1',\dots,a_n')$}
\end{align*}

This motivates us to think of the above extensionality property as an operation
which, given terms which relate elements in the context, substitutes these connections into $t$ to get a relation between the corresponding instances of $t$.

In this interpretation, the proof that a closed term $t$ is related to itself
\[\mathsf{r}(t) : t R_T t\]
has specific computational content.  Furthermore, the algorithm
associated to this proof has the same structure as $t$ itself.

\section{The dependent case} \label{s:dep-ext}
To make matters simple, we use PTS formulation of dependent type theory with ``type-in-type''.  This system is denoted as $\lstar$.  It has a universal type $\sta$, the type of all types.  This allows us to unify into one the three classical judgement forms of dependent type theory:
\begin{align*}
&\Gamma \vdash A\ \; \btype\\
&\Gamma \vdash a:A\\
&\Gamma \vdash B : (A)\btype
\end{align*}
The judgment $\Gamma \vdash A\ \; \btype$ is replaced by $\Gamma \vdash A : \sta$.  Similarly, $\Gamma \vdash (A)B\ \; \btype$ is replaced by $\Gamma, x{:}A \vdash B : \sta$.  Thus types and terms of type $\sta$ are completly identified.

The syntax of $\lstars$, the type-in-type PTS with $\Sigma$-types\footnote{We call the system a ``PTS'' because its notion of equality is based on
untyped conversion of lambda terms.  Classically, $\Sigma$ types are
not part of the PTS formalism, but including them here
presents no difficulty.} is
\begin{align*}
t ::= \sta &\mid x \mid \Pi x {:} A. B(x) \mid \Sigma x{:}A. B(x)\\
&\mid \lambda x{:}A.t(x) \mid s t
\mid (s,t) \mid \pi_1 t \mid \pi_2 t
\end{align*}

A notational note:\ the parentheses following the matrix of the $\Pi$, $\Sigma$, and $\lambda$ constructors are not part of the syntax, and merely pronounce the fact that the term may depend on the variables in question.  In general, when we write $t = t(x_1,\dots,x_n)$, we do not commit to having displayed all the free variables of $t$; it is never mandatory to display a free variable.

The purpose of this notation is merely to reduce clutter in anticipation of substitution of $t$ by an instance of (some of the) variables.  Our general notation for substituting a free variable $x$ in $t$ by $a$ is
\[t[a/x]\]
In particular, if $t = t(x_1,\dots,x_n)$, then
\[ t[a_1/x_1]\cdots[a_n/x_n] = t(a_1,\dots,a_n)\]

In the following development, we shall consider the open term model of the above type theory, using the same theory as our meta-level.  To simplify notation, we write $\semof{A}$ simply as $A$.  As well, if $t(x_1,\dots,x_n) : T(x_1,\dots,x_n)$, then
$\semof{t}_{x_1,\dots,x_n:=a_1,\dots,a_n}$ is denoted as $t(a_1,\dots,a_n)$.

The only axiom of this type system has the form $\sta : \sta$, asserting that the universe of types $\sta$ is itself a type.  Its intuitive meaning is
\begin{quotation}
The collection of structures which types are interpreted by forms the same kind of structure.
\end{quotation}

In particular, if types are interpreted by types-with-relation $R_A : A \to A \to \sta$, then this interpretation must also include a relation on the universe of types
\[ R_\sta : \sta \to \sta \to \sta \]

But how should this relation interact with objects inhabiting related types?

To answer this question, let us consider how the previous extensionality theorem could be extended to the dependent case.  Suppose we are given terms
\begin{align}
x {:} A &\vdash B(x) : \sta \label{eq1}\\
x {:} A &\vdash b(x) : B(x)
\end{align}
If we are now given $a:A$, $a':A$, we want to conclude that
\begin{equation}R_A a a' \to R_B b(a) b(a') \label{eq2}
\end{equation}
However, the two terms $b(a)$ and $b(a')$ have different types!  We need additional structure to formulate extensionality of dependent maps.

Looking again at \eqref{eq1}, observe that, by extensionality, from any witness $a^*$ of the hypothesis of \eqref{eq2}, it should be possible to construct a witness $B(a^*)$ to the relation $R_\sta B(a) B(a')$.  The relation between $b(a)$ and $b(a')$ can then be construed as lying over the relation between $B(a)$ and $B(a')$.  This suggests the following principle:

\begin{quotation}
  Every witness $E : R_\sta A B$ to the fact that $A$ and $B$ are related elements of the universe induces a relation
\[\tilde E : A \to B \to \sta\]
between elements of corresponding types.
\end{quotation}

Let $e : R_\sta A B$ be given.  In the sequel, we often write $\tilde e a b$ as $a \sim_e b$.  This relation should have the following properties:
\begin{itemize}
\item For any $a:A$, there exists an element $e(a) : B$ which is
$R_B$-minimal with respect to all elements $b$ which are $\sim_e$-related to from $a$; thus
\begin{itemize}
\item   $a \sim_e e(a)$
\item   $a \sim_e b \RA e(a) R_B b$
\end{itemize}
\item 
For any $b : B$, there exists an element $\bar e(b) : A$ which is
$R_A$-maximal with respect to all elements $a$ which are $\sim_e$-related to $b$:
\begin{itemize}
\item $\bar e(b) \sim_e b$
\item $a \sim_e b \RA a R_A \bar e(b)$
\end{itemize}
\item 
Whenever $e(a)$ is related to $b$, $a$ is related to $b$.  Whenever $a$ is related to $\bar e(b)$, $a$ is related to $b$.  Thus the two implications above are invertible.
\end{itemize}

(We remark that, if the relations in question are assumed to be reflexive and transitive, then the above conditions on $\tilde e$ define a connection between $A$ and $B$ in the sense of order theory:
\begin{align*}
\lambda x{:}A. e(x) &: A \to B\\
\lambda y{:}B. \bar e(y) &: B \to A\\
\forall x{:}A \forall y{:}B \quad &x R_A \bar{e}(y) \iff e(x) R_B y.\qquad)
\end{align*}

So --- if $x{:}A \vdash B(x):\sta$, and $a^* : aR_Aa'$, we have
$B(a^*)^\sim : B(a) \to B(a') \to \sta$.

We now say that a family of relations
$\setof{R_A : A \to A \to \sta \mid A : \sta}$ is \emph{logical} if:
\begin{align}
R_{\Pi x{:}A.B(x)} f f' &= \prod_{a{:}A} \prod_{a'{:}A} \Pi a^*: R_A a a'.
B(a^*)^\sim (f a) (f' a')\\
R_{\Sigma x{:}A.B(x)} p p' &= \Sigma a^* : R_A\, (\pi_1 p)\, (\pi_1 p'). B(a^*)^\sim \,
(\pi_2 p)\, (\pi_2 p')
\end{align}

The general statement of extensionality will take the following form:

\begin{thm}
  Let $\setof{R_A}$ be logical.  For every term $t$ typed in the context
\[ x_1 : A_1, \cdots, x_n:A_n(x_1,\dots,x_{n-1}) \vdash t(x_1,\dots,x_n)
:T (x_1,\dots,x_n)\]
and for any pair of coordinate-wise related instances
\begin{align*}
&a_1 : A_1, \cdots, a_n : A_n(a_1,\dots,a_{n-1})\\
&a_1' :A_1, \cdots, a_n': A_n(a_1',\dots,a_{n-1}')\\
&a_1^* : R_{A_1} a_1 a_1', \cdots, a_n^* : A_n(a_1^*,\dots,a_{n-1}^*)^\sim\; a_n a_n'
\end{align*}
there is a witness $t(a_1^*,\dots,a_n^*)$ to the fact that
\[ t(a_1,\dots,a_n) \sim_{T(a_1^*,\dots,a_n^*)} t(a'_1,\dots,a'_n).\]
\end{thm}

\comment{The conclusion of the theorem may be alternatively displayed as
\[t [\vec {a^*} / \vec x] : t [\vec a/\vec x] \sim_{T [\vec {a^*}/\vec x]}
t [\vec {a'}/\vec x] \]}

In particular, when $t:T$ is a closed term, the above principle yields
a new term $t() : t \sim_{T()} t$ which is not quite the same as $t$,
because it is one dimension higher.
Let us instead write $t^*$ for this term.
For such closed terms, we get

\[t^* : t \sim_{T^*} t\]
\[T^* : T \sim_{\sta^*} T\]
\[ \sta^* : \sta \sim_{\sta^*} \sta \]

This suggests an answer to the quesion of what should be the logical condition on the universe constant $\sta$.  We should have that
\[\forall A : \sta\quad a R_A a' \iff a \sim_{A^*} a'\]
In particular,
\[ A R_\sta B \iff A \sim_{{\sta}^*} B\]

The problem with the statement of dependent extensionality theorem above is that the formula in the conclusion already makes reference to the result
of the substitution of $a_1^*,\dots,a_n^*$ into $T(\vec x)$, which requires the extensionality of the judgement
$\Gamma \vdash T(\vec x) : *$ to be known beforehand.  In general, the proof of this fact will again depend on
extensionality of subterms appearing in $T$.

We therefore move to first represent the type universe of $\lstar$ in a
minimal extension of the system relevant for this purpose.  The above
theorem will be stated for the result of reflecting the meta-level into this
universe.  The next step is to mutually define the type of equivalences between
two elements of this universe, and the corresponding relations induced
by such equivalences.  The inter-dependency between these concepts is
resolved using an indexed inductive--recursive definition
of \cite{Dybjer2001}, and this
allows us to state the above theorem for the (reflected) universe.
Finally, we prove the theorem by induction on the structure of derivations.

\section{$\lstars$ in $\lstars$}

The inductive--recursive definition of the universe $U$ of $\lstars$-types is as follows:
\begin{align*}
\mathsf{Inductive}\ &U : * \;\; :=\\
&|\ \qFun : \Pi A:U. (TA \to U) \to U\\
&|\ \qSum : \Pi A:U. (TA \to U) \to U\\
&|\ \qsta : U\\
\mathsf{with} \quad &T : U \to * \;\; :=\\
&T(\qFun A B) = \Pi a:TA. T[Ba]\\
&T(\qSum A B) = \Sigma a:TA. T[Ba]\\
&T(\qsta) = U
\end{align*}

Let $\lstau$ be $\lstars$ augmented with the above datatype.  Notice that every derivation
in $\lstars$ is also a derivation in $\lstau$.  \comment{We shall write $\Gamma \vdash M : A$ to denote
a derivation in $\lstar$ and $\Gamma \vdash_U M : A$ to denote a derivation in $\lstau$.}
\begin{dfn}
  We define a map $\ol{(\cdot)}$ from the raw terms of $\lstar$ to the raw terms of $\lstau$ as follows:
  \begin{align*}
    \ol \sta &= \qsta\\
    \ol x &= x\\
    \ol {\Pi x{:}A.B} &= \qFun \ol A (\lambda x{:}T \ol A. \ol B)\\
    \ol {\Sigma x{:}A.B} &= \qSum \ol A (\lambda x{:}T \ol A. \ol B)\\
\comment{    \ol {\lambda x{:}A.t} &= \qlam (\lambda x{:}\ol A. \ol t)\\
    \ol {s t} &= \qapp \ol s \ol t\\
    \ol {(s,t)} &= \qpair \ol s \ol t\\
    \ol {\pi_1 p} &= \qproj1 \ol p\\
    \ol {\pi_2 p} &= \qproj2 \ol p}
  \ol {\lambda x{:}A.t} &= \lambda x{:}\ol A.\ol t\\
  \ol {s t} &= \ol s \ol t\\
  \ol {(s,t)} &= (\ol s, \ol t)\\
  \ol {\pi_i s} &= \pi_i \ol s
\end{align*}
\end{dfn}

\begin{dfn}
Suppose $\Gamma = \setof{x_1 : A_1, \dots, x_n : A_n}$.
We define \[ \ol \Gamma := \setof{x_1 : T \ol A_1, \dots, x_n : T \ol{A_n}}\]
\end{dfn}

\begin{lemma}
  (Substitution Lemma) \label{subst}
  Let $M,N$ be $\lstar$-terms.  Then
\[  \ol{M[N/x]} = \ol M [\ol N/x] \]
\end{lemma}

\begin{cor} \label{conv}
  Let $M = N$ be $\lstar$-terms. Then 
  \[ M = N \quad \then \quad \ol M = \ol N\]
\end{cor}

\begin{thm}
  (Reflection of $\sta$ into $U$)
\comment{  For every $\lstars$-context $\Gamma$, pseudoterms $M$ and $A$, and
a $\lstars$-derivation $\delta$ of $\Gamma \vdash M : A$, there exists a $\lstau$-context
$\ol \Gamma$, term $\ol A : U$, pseudoterm $\ol M$, and a derivation $\ol \delta$ of
$\ol \Gamma \vdash \ol M : T \ol A$.  Schematically,}
\[ \Gamma \vdash_{\lstars} M : A \then \ol \Gamma \vdash_{\lstau} \ol M : T \ol A\]
  
\end{thm}

\begin{proof}
The translation is done by induction on $\Gamma \vdash M : A$.
\begin{description}
\item[Axiom] $\vdash_\lstarss \sta : \sta$.  Then $\ol \Gamma = \Gamma = ()$.  Also $\ol A = \ol M = \qsta$.

The conversion rule gives
\begin{prooftree}
  \AXC{$ \qsta : U$}
  \AXC{$ T \qsta : \sta$}
  \AXC{$ U = T \qsta$}
  \TIC{$ \qsta : T \qsta$}
\end{prooftree}
Thus indeed $\vdash_U \ol M : T \ol A$.
\item[Variable]  Suppose $\delta$ ends with
  \begin{prooftree}
  \AXC{$\Gamma \vdash A : \sta$}
  \UIC{$\Gamma, x:A \vdash x:A$}
  \end{prooftree}
  By induction hypothesis, we have
  \[ \ol \Gamma \vdash \ol A : T \ol{\qsta}\]
  Hence $\ol \Gamma \vdash \ol A : U$, and $\ol \Gamma \vdash T \ol A : \sta$
  By the variable rule, we have
  \[ \Gamma, x : T \ol A \vdash x : T \ol A \]
\item[Weakening] Let the derivation end with
  \begin{prooftree}
  \AXC{$\Gamma \vdash M : A$}
  \AXC{$\Gamma \vdash B : \sta$}
  \BIC{$\Gamma, y:B \vdash M:A$}
\end{prooftree}
By induction hypothesis, we have
\begin{align*}
  \ol \Gamma &\vdash \ol M : T \ol A\\
  \ol \Gamma &\vdash \ol B : T \ol \qsta
\end{align*}
That is, $\ol \Gamma$ yields $\vdash \ol B : U$.  Then $T \ol B : \sta$.
By weakening,
\[ \ol \Gamma, y : T \ol B \vdash \ol M : T \ol A \]
\item[$\Pi$-formation] 
Given
\begin{prooftree}
  \AXC{$\Gamma \vdash A : \sta$}
  \AXC{$\Gamma, x : A \vdash B : \sta$}
  \BIC{$\Gamma \vdash \Pi x{:}A.B : \sta$}
\end{prooftree}
the induction hypotheses yield
\begin{align}
  \ol \Gamma &\vdash \ol A : T \ol \sta    \label{piform1}\\
  \ol \Gamma, x : T \ol A &\vdash \ol B : T \ol \sta \label{piform2}
\end{align}
Since $\ol A, \ol B : T \ol \sta = T \qsta = U$,
we have $\ol \Gamma \vdash T \ol A : \sta$ as well
as $\Gamma, x : T\ol A \vdash  T\ol B : \sta$.

By the $\Pi$-introduction rule, \eqref{piform2} yields
\[\ol \Gamma \vdash \lambda x {:} T \ol A. \ol B  : T \ol A \to U\]
whence $\Pi$-elimination together with \eqref{piform1} yields
\[\ol \Gamma \vdash \qFun \ol A (\lambda x{:}T\ol A. \ol B) : U \]
That is,
\[ \ol \Gamma \vdash \ol {\Pi x{:}A. B} : T \ol \sta \]
\item[$\Sigma$-formation] Treated in an analogous fashion.
\item[$\Pi$-introduction] Suppose the derivation is of the form
  \begin{prooftree}
    \AXC{$\Gamma \vdash A : *$}
    \AXC{$\Gamma, x : A \vdash B : *$}
    \AXC{$\Gamma, x : A \vdash b : B$}
    \TIC{$\Gamma \vdash \lambda x{:}A.b : \Pi x{:}A. B$}
  \end{prooftree}
The induction hypotheses give us
\begin{align}
  \ol \Gamma &\vdash \ol A : T \ol \sta   \notag \\
  \ol \Gamma, x : T \ol A &\vdash \ol B : T \ol \sta \notag \\
  \ol \Gamma, x : T \ol A &\vdash \ol b : T \ol B \label{piintro}
\end{align}
As in the previous case, we actually have
\begin{align*}
    \ol \Gamma &\vdash \ol A : U & \ol \Gamma &\vdash T \ol A : *\\
  \ol \Gamma, x : T \ol A &\vdash \ol B : U
  &\ol \Gamma, x : T \ol A &\vdash T \ol B : *
  \\
 \ol \Gamma &\vdash \ol {\Pi x{:}A. B} : U
 &\ol \Gamma &\vdash T [\ol {\Pi x{:}A. B}] : *
\end{align*}
By $\Pi$-introduction
on \eqref{piintro}, we have
\[\ol \Gamma \vdash \lambda x {:} T \ol A. \ol b : \Pi x{:}T \ol A. T \ol B\]
But we also find that \begin{equation} \Pi x{:}T \ol A. T \ol B =
T [\qFun \ol A (\lambda x{:} T \ol A. \ol B)] = T \ol {\Pi x{:}A. B}
\label{TolPi}
\end{equation}
and so conclude that
\[ \ol \Gamma \vdash \ol {\lambda x{:}A.b} : T \ol {\Pi x{:}A. B}\]

\item[$\Pi$-elimination] Suppose we are given
  \begin{prooftree}
    \AXC{$\Gamma \vdash A : *\quad \Gamma, x: A \vdash B : *$}
    \AXC{$\Gamma \vdash f : \Pi x{:}A.B$}
    \AXC{$\Gamma \vdash a : A$}
    \TIC{$\Gamma \vdash f a : B[a/x]$}
  \end{prooftree}
  The induction hypothesis yield, on the one hand, that
\begin{align*}
    \ol \Gamma &\vdash \ol A : U & \ol \Gamma &\vdash T \ol A : *\\
  \ol \Gamma, x : T \ol A &\vdash \ol B : U
  &\ol \Gamma, x : T \ol A &\vdash T \ol B : *
  \\
 \ol \Gamma &\vdash \ol {\Pi x{:}A. B} : U
 &\ol \Gamma &\vdash T [\ol {\Pi x{:}A. B}] : *
\end{align*}
and on the other hand, that
\begin{align*}
  \ol \Gamma &\vdash \ol f : T \ol{\Pi x{:}A.B}\\
  \ol \Gamma &\vdash \ol a : T \ol A
\end{align*}
Since $\ol f$ by conversion in \eqref{TolPi} has type $\Pi x{:}T \ol A. T \ol B$,
we may write
\[\ol \Gamma \vdash \ol f \ol a : T \ol B[\ol a / x] \]
By Lemma \ref{subst}, the type in the above judgment is equal to $T\ol{B[a/x]}$
.
\item[$\Sigma$-introduction]
When we are at
\begin{prooftree}
  \AXC{$\Gamma \vdash A : * \quad \Gamma, x:A \vdash B : *$}
  \AXC{$\Gamma \vdash a : A$}
  \AXC{$\Gamma \vdash b : B[a/x]$}
  \TIC{$\Gamma \vdash (a,b) : \Sigma x{:}A.B$}
\end{prooftree}
the induction hypotheses give as before that
\begin{align*}
\ol \Gamma &\vdash T \ol A : \sta \\
\ol \Gamma, x : T \ol A &\vdash T \ol B : \sta
\end{align*}
and, in addition, we also have
\begin{align*}
  \ol \Gamma &\vdash \ol a : T \ol A\\
  \ol \Gamma &\vdash \ol b : T \ol{B[a/x]}
\end{align*}
Recall that
\begin{align*}
\ol{\Sigma x{:}A.B} &= \qSum \ol A (\lambda x{:}T\ol A. \ol{B})\\
T \ol {\Sigma x{:}A.B} &= \Sigma x{:}T \ol A. T \ol B
\end{align*}

By Lemma \ref{subst}, $T \ol {B[a/x]} = T \ol B [\ol a /x]$.  Hence
$b : T \ol B [\ol a /x]$.

By $\Sigma$-introduction, we now obtain
\[\ol \Gamma \vdash (\ol a, \ol b) : \Sigma x{:}T \ol A. T \ol B\]
In other words, $\ol \Gamma \vdash \ol{(a,b)} : T \ol {\Sigma x{:}A.B}$.
\item[$\Sigma$-elimination] Let there be derived
  \begin{prooftree}
    \AXC{$\Gamma \vdash A : \sta$}
    \AXC{$\Gamma, x:A \vdash B : \sta$}
    \AXC{$\Gamma \vdash p : \Sigma x{:}A.B$}
    \TIC{$\Gamma \vdash \pi_1 p : A$}
    \noLine
    \UIC{$\Gamma \vdash \pi_2 p : B[\pi_1 p / x]$}
  \end{prooftree}
  Assume we have
  \begin{align*}
    \ol \Gamma &\vdash \ol A : U\\
    \ol \Gamma, x : T \ol A &\vdash \ol B : U\\
    \ol \Gamma &\vdash \ol p : T \ol {\Sigma x{:}A.B}
  \end{align*}
  We have just seen that
  $T \ol {\Sigma x{:}A.B} = \Sigma x{:}T \ol A. T \ol B$.
  Thus
\[ \pi_1 \ol p : T \ol A\]
\[ \pi_2 \ol p : [T \ol B][\pi_1 \ol p / x]\]

The subjects of these judgements can be rewritten as $\ol{\pi_i p}$.

Also
\[ [T \ol B] [\pi_1 \ol p / x] = [T \ol B] [\ol {\pi_1 p} / x]  = T [\ol B
[\ol {\pi_1 p}/x]] = T [\ol {B [{\pi_1 p} / x]}]\]
Thus we have
\begin{align*}
  \ol \Gamma &\vdash \ol {\pi_1 p} : T \ol A\\
  \ol \Gamma &\vdash \ol {\pi_2 p} : T \ol {B[{\pi_1 p} /x]}
\end{align*}
\item[Conversion] 
Suppose we come across
\begin{prooftree}
  \AXC{$\Gamma \vdash M : A$}
  \AXC{$\Gamma \vdash B : *$}
  \AXC{$A = B$}
  \TIC{$\Gamma \vdash M : B$}
\end{prooftree}
By induction hypothesis, we have
\begin{align*}
  \ol \Gamma &\vdash \ol M : T \ol A\\
  \ol \Gamma &\vdash \ol B : U
\end{align*}
By Lemma \ref{conv}, we have
\[ \ol A = \ol B \]
But clearly that implies that
\[ T \ol A = T \ol B \]
It is likewise clear that
\[ \ol \Gamma \vdash T \ol B : * \]
By the conversion rule, we comprehend
\[\ol \Gamma \vdash \ol M : T \ol B \]
\end{description}
This completes the proof of the theorem.
\end{proof}

\section{Extensionality of $\lstar$}

We work in $\lstau$.  Assume as given a relation
\[ R : \Pi A : U. T A \to T A \to \sta \]
At this point we begin to denote $R A$ by the symbol
\[ A^\eqv : T A \to T A \to \sta \]
and we shall often write $A^\eqv a a'$ as $a \ee{A} a'$.

We also require that every inhabitant $e$ of
$\qsta^\eqv A B$ gives rise to a relation
\[e^\sim : T A \to T B \to \sta \]
and we often write $e^\sim a b$ as $a \sim_e b$.

For the notion of equivalence of types, we assume as
given a binary relation on the type $U$:
\[\ee{} : U \to U \to U\]
and we write $A {\ee{}} B$ for ${\ee{}} A B$.  This notation is consistent,
because we shall stipulate that
\[ A \ee{} B \quad=\quad A \ee{\sta}\! B \]

In order to precisely state extensionality theorem using the
above data, we must provide answers to the following
questions:
\begin{itemize}
\item What does it mean for two types $A, B : U$ to be equivalent?
\item What does it mean for two elements to be related by an
equivalence?
\end{itemize}

To answer these questions, we proceed as in Section
\ref{s:dep-ext}.
We identify the notion of equivalence with the notion
of a binary relation with certain properties.  Semantically, $e$ is
an equivalence between $A$ and $B$ if $e \subseteq \semof{A} \times 
\semof{B}$, and $e$ satisfies those additional properties.

In order to represent equivalences in type theory, we must therefore
introduce a syntax for defining such binary relations between two
types.

That is, for any two terms $A,B : U$, we must introduce a type of
\emph{codes of equivalences from $A$ to $B$}.  This type will
be denoted as
\[ \eqtype(A,B) \]
Simulateneously with this type, we must also define a function which
evaluates the codes to actual relations between the types $A$ and $B$.
Ie, we need a map
\[ \reltype : \eqtype(A,B) \to TA \to TB \to \sta \]

This suggests that the type constructors $\ee{}$ and $\sim_e$ can
be captured using a variant of the inductive--recursive (IR) definitions.

Upon reflecting on this possibility,
it shall become manifest that the two concepts above cannot be defined
uniformly in $A$ and $B$; rather, the $A$ and $B$ must
take part in the recursive construction of both the set
$\eqtype(A,B)$ as well as the map $\reltype$.
Thus, the arguments $A$ and $B$ are to be treated as
\emph{indices}, so that we are dealing with
an \emph{indexed inductive--recursive definition} (IIRD). 

We are now in the position to answer the two questions posed above.
The notion of equivalence of types $A$ and $B$ and the notion of
elements of the corresponding types being related over this
equivalence
are both defined simultaneously by indexed induction--recursion.
The definition follows.
\begin{align*}
\mathsf{Inductive}\ &\eqtype : U \to U \to * \;\; :=\\
&|\ \eqsta : \eqtype \qsta \qsta\\
&|\ \eqfun \setof{A A' : U} \setof{B : A \to U} \setof{B' : A' 
\to U}\\
&\quad (A^* : \eqtype A A')
(B^* : \Pi a{:}A \Pi a'{:} A' \Pi a^* : \reltype A^* a a'.
B a \ee{} B' a')\\
&\qquad: \eqtype (\qFun A B) (\qFun A' B')\\
&|\ \eqsum \setof{A A' : U} \setof{B : A \to U} \setof{B' : A' 
\to U} \\
&\quad(A^* : \eqtype A A')
(B^* : \Pi a{:}A \Pi a'{:} A' \Pi a^* : \reltype A^* a a'.
B a \ee{} B' a')\\
&\qquad: \eqtype (\qSum A B) (\qSum A' B')\\
\mathsf{with} \quad &\reltype : \Pi \setof{A} \{B\} : U.\;
\eqtype A B \to T A \to TB \to * \\
&\reltype (\eqsta) A B = \eqtype A B\\
&\reltype (\eqfun A^* B^*) f f' = \Pi x{:}A \Pi x'{:}A' \Pi x^* :
\reltype A^* x x'. \\
&\quad
  \reltype (B^* x x' x^*) (f x) (f' x')\\
  &\reltype (\eqsum A^* B^*) p p' =
  \Sigma x^* : \reltype A^* (\pi_1 p) (\pi_1 p'). \\
&\quad
  \reltype (B^* (\pi_1 p) (\pi_1 p') x^*)
  (\pi_2 p) (\pi_2 p')
\end{align*}

We denote the system $\lstau$ extended with the above IIRD by
$\lstaue$.

We remark that $\lstau$ is a subsystem of $\lstaue$ in the sense that
every term of $\lstau$ is a term of $\lstaue$, and every derivation
in $\lstau$ is also a derivation in $\lstaue$.

\begin{dfn}
  We define two operations on those terms of $\lstau$ which are in the
image of the reflection map $\ol{\cdot} : \lstar \to \lstau$.
  
  Thus, the operations are really defined on terms of form $\ol M$ or $(T)\ol A$,
but for notational convenience we shall write these as $M$ and $A$ just as well.

  The first operation marks every variable with an apostrophe:
\[ (-)' : \mathsf{Terms}(\lstau) \to \mathsf{Terms}(\lstau) \]
\begin{align*}
  \qsta' &= \qsta\\
  (x)' &= x'\\
  (\qFun A B)' &= \qFun A' B'\\
  (\qSum A B)' &= \qSum A' B'\\
  (\lambda x{:}A.b)' &= \lambda x'{:}A'.b'\\
  (st)' &= s't'\\
  (s,t)' &= (s',t')\\
  (\pi_1 p)' &= \pi_1 p'\\
  (\pi_2 p)' &= \pi_2 p'
\end{align*}

The second operation substitutes every type by an
equivalence and every term by a higher-dimensional cell.
\[ (-)^* : \mathsf{Terms}(\lstau) \to \mathsf{Terms}(\lstaue) \]
\begin{align*}
  (x)^* &= x^*\\
  \qsta^* &= \eqsta\\
  (\qFun A B)^* &= \eqfun A^* (\lambda x{:}A\; \lambda x'{:}A'\;
  \lambda x^* : \reltype{A^*} x x'.   B^*)\\
  (\qSum A B)^* &= \eqsum A^* (\lambda x{:}A\; \lambda x'{:}A'\;
  \lambda x^* : \reltype{A^*} x x'.   B^*)\\
  (\lambda x{:}A.b)^* &= \lambda x{:}A\; \lambda x'{:}A'\;
  \lambda x^* : \reltype {A^*} x x'.   b^*\\
  (f a)^* &= f^* a a' a^*\\
  (a,b)^* &= (a^*,b^*)\\
  (\pi_1 p)^* &= \pi_1 p^*\\
  (\pi_2 p)^* &= \pi_2 p^*
\end{align*}
\end{dfn}

\begin{thm}
$(M[N/x])' = M'[N'/x']$
\end{thm}
\begin{thm}
$(M[N/x])^* = M^*[N/x,N'/x',N^*/x^*]$
\end{thm}
\begin{proof}
  \begin{description}
  \item[Axiom] $(\qsta [N/x])^* = (\qsta)^* = \eqsta = \eqsta [N/x,N'/x',N^*/x^*] $
  \item[Variable] \[(y[N/x])^* = \begin{cases}
(x[N/x])^* = N^* = x^* [N/x,N'/x',N^*/x^*] \qquad&y=x\\
(y[N/x])^* = y^* = y^* [N/x,N'/x',N^*/x^*] &y \neq x
\end{cases}\]
\item[Product]
\begin{align*}
 (\qFun A B\; [N/x])^* &= (\qFun A[N/x] B[N/x])^*\\
&= \eqfun (A[N/x])^* (B[N/x])^* \\
&= \eqfun A^*[N/x,N'/x',N^*/x^*] B^*[N/x,N'/x',N^*/x^*]\\
&= (\eqfun A^* B^*) [N/x,N'/x',N^*/x^*]
\end{align*}
\item[Sum]
\begin{align*}
 (\qSum A B\; [N/x])^* &= (\qSum A[N/x] B[N/x])^*\\
&= \eqsum (A[N/x])^* (B[N/x])^* \\
&= \eqsum A^*[N/x,N'/x',N^*/x^*] B^*[N/x,N'/x',N^*/x^*]\\
&= (\eqsum A^* B^*) [N/x,N'/x',N^*/x^*]
\end{align*}
\item[Abstraction] We remark that the Variable Convention can has observed.
  \begin{align*}
    ((\lambda y{:}A.b)[N/x])^* &= (\lambda y:A[N/x].b[N/x])^*\\
&= \lambda y{:}A[N/x]\; \lambda y'{:}(A[N/x])'\; \lambda y^* :
\reltype (A[N/x])^* y y'. (b[N/x])^*\\
&= \lambda y{:}A[N/x]\;\lambda y'{:}A'[N'/x']\;
\lambda y^* : \reltype A^*[N/x,N'/x',N^*/x^*] y y'.\\
&\hspace{6cm}b^* [N/x,N'/x',N^*/x^*]\\
&= \lambda y{:}A[N,N',N^*/x,x',x^*]\;
\lambda y'{:}A'[N,N',N^*/x,x',x^*]\;\\
&\hspace{1cm}\lambda y^*:(\reltype A^* y y') [N,N',N^*/x,x',x^*]. 
b^* [N,N',N^*/x,x',x^*]\\
&= (\lambda y{:}A\; \lambda y'{:}A'\; \lambda y^* : \reltype A^* y
y'. b^*) [N,N',N^*/x,x',x^*]\\
&= (\lambda y : A. b)^*[N/x,N'/x',N^*/x^*]
  \end{align*}
\item[Application] 
  \begin{align*}
    (s t [N/x])^* &=(s [N/x] t [N/x])^*\\
&= (s [N/x])^* (t [N/x]) (t [N/x])' (t [N/x])^*\\
&= (s^* [N,N',N^*/x,x',x^*] t[N/x] t'[N'/x] t^* [N,N',N^*/x,x',x^*]\\
&= (s^* t t' t^*)[N,N',N^*/x,x',x^*]\\
&= (s t)^* [N,N',N^*/x,x',x^*]
  \end{align*}
\item[Pairing] 
  \begin{align*}
    ((s,t)[N/x])^* &= (s[N/x],t[N/x])^*\\
    &= ( (s[N/x])^*, (t[N/x])^*)\\
    &= ( s^*[N/x,N'/x',N^*/x^*], t^*[N/x,N'/x',N^*/x^*])\\
    &= (s^*,t^*)[N,N',N^*/x,x',x^*]\\
    &= (s,t)^*[N,N',N^*/x,x',x^*]
  \end{align*}
\item[Projection] 
  \begin{align*}
    ((\pi_i t) [N/x])^* &= (\pi_i t [N/x])^*\\
&= \pi_i (t [N/x])^*\\
&= \pi_i (t^* [N,N',N^*/x,x',x^*])\\
&= \pi_i t^* [N,N',N^*/x,x',x^*]\\
&= (\pi_i t)^* [N,N',N^*/x,x',x^*]
\qedhere
  \end{align*}
\end{description}
\end{proof}

\begin{cor} \label{c:conv}
  Suppose $M=N$.  Then $M^*=N^*$.
\end{cor}
\begin{proof}
Assume $M = (\lambda x{:}A.s)t$ and $N = s[t/x]$.   We have
  \begin{align*}
    M^* = ((\lambda x{:}A.s)t)^* &= (\lambda x{:}A.s)^* t t' t^*\\
&= (\lambda x{:}A\; \lambda x'{:}A'\; \lambda x^* : \reltype A^* x
x'. s^*) t t' t^*\\
&= s^*[t/x][t'/x'][t^*/x^*]\\
&= (s[t/x])^* = N^*
  \end{align*}
where the last equality is by the previous proposition.

Now suppose that $M = \pi_i (t_1,t_2)$, and $N = t_i$.
Then \[ M^* = \pi_i(t_1^*,t_2^*) = t_i^* = N^* \qedhere\]
\end{proof}

\begin{dfn}
  A $\lstau$-context $\Gamma$ is said to be a \emph{$U$-context} if
$\Gamma$ is of the form
\[x_1 : T A_1, \dots, x_n : T A_n(x_1,\dots,x_{n-1}) \]
and for $0 \le i < n$, it holds that
\[x_1:T A_1, \dots, x_i : T A_i(x_1,\dots,x_{i-1}) \vdash
A_{i+1}(x_1,\dots,x_i) : U\]

If $\Gamma$ is a $U$-context, and $\Gamma \vdash A : U$, we call
$A$ a \emph{$U$-type in $\Gamma$}.
\end{dfn}

\begin{dfn}
Given a $U$-context $\Gamma = \setof{x_1 {:} T A_1, \dots x_n{:}T A_n}$, put
\[
\Gamma^* = \begin{cases}
x_1 : T A_1,\ \dots,\ x_n : T A_n,\\
x_1' : T A_1',\ \dots,\ x_n' : T A_n',\\
x_1^* : \reltype A_1^* x_1 x_1',\ \dots,\ x_n^* : \reltype A_n^* x_n
x_n'
\end{cases} \]
\end{dfn}

Let $\Gamma'$ be obtained from $\Gamma$ by apostrophizing every
variable, including those occurring in their declared types.
Obviously, we can have
\begin{thm}
  $\Gamma \vdash M : A \then \Gamma' : M' : A'$.
\end{thm}

\begin{thm}
  Let $\Gamma$ be a $U$-context, and $A$ a $U$-type in $\Gamma$.
  Then
\begin{equation}
 \Gamma \vdash_{\lstau} M : T A \quad \then \quad
\Gamma^* \vdash_{\lstaue} M^* : \reltype A^* M M'
\label{e:ext}
\end{equation}
\end{thm}
\begin{proof}
  We proceed by induction on the derivation.
  \begin{description}
  \item[Axiom] Suppose $\Gamma \vdash \qsta : T \qsta$.  We have
\[ \qsta^* = \eqsta : \eqtype \qsta \qsta = \reltype \eqsta \qsta
\qsta = \reltype \qsta^* \qsta \qsta'\]
where $\eqsta : \eqtype \qsta \qsta$ in any context.

By conversion rule, $\Gamma^* \vdash \qsta^* : \reltype \qsta^* \qsta \qsta'$.
\item[Variable] Suppose we have a derivation tree with root
  \begin{prooftree}
 \AXC{$\Gamma \vdash A : T \qsta$}
 \UIC{$\Gamma, x : TA \vdash x : T A$}
  \end{prooftree}
(Notice that the hypothesis says that $A$ is a $U$-type in $\Gamma$.)

By the previous proposition, $\Gamma' \vdash A' : T \qsta$.

By induction hypothesis, $\Gamma^* \vdash A^* : \reltype \qsta^* A
A'$.

Since $\reltype \qsta^* A A' = \eqtype A A'$, we have 
$\Gamma^* \vdash A^* : \eqtype A A'$ by conversion.

Yet $\Gamma^*$ also yields that $T A : *$ and $T A' : *$, and thus we
may form the context $\Gamma^*, x:TA, x':TA' \vdash $.  In this context,
we may derive that
\[\Gamma^*, x:TA, x':TA' \vdash \reltype A^* x x' : *\]
using the typing rule for the $\reltype$ constructor.

By the variable rule, we have
\[ \Gamma^*, x:TA, x':TA',x^*:\reltype A^* x x' \vdash x^* :
\reltype A^* x x'\]

The context in the above judgement is $(\Gamma, x:TA)^*$.  The subject is
$(x)^*$.  The type predicate is as displayed in \eqref{e:ext}. 
\item[Weakening]
Suppose they give you
\begin{prooftree}
  \AXC{$\Gamma \vdash M : T A$}
  \AXC{$\Gamma \vdash B : T \qsta$}
  \BIC{$\Gamma, y : T B \vdash M : T A$}
\end{prooftree}

The induction hypotheses give that
\begin{align*}
  \Gamma^* &\vdash M^* : \reltype A^* M M'\\
  \Gamma^* &\vdash B^* : \reltype \qsta^* B B'
\end{align*}

As before, we may conclude that $B, B' : U$ in $\Gamma^*$, that $B^* : \eqtype B B'$,
and that $\Gamma^*, y : T B, y' : T B'$ is a valid context.

Then $\reltype B^* y y' : *$, and by weakening we get
\[ (\Gamma, y : T B)^* \vdash M^* : \reltype A^* M M' \]
\item[Formation] Consider the typing
\begin{prooftree}
\AXC{$\Gamma \vdash A : T \qsta$}
\AXC{$\Gamma, x {:} TA \vdash B : T \qsta$}
\BIC{$\Gamma \vdash \qFun A (\lambda x{:} T A. B) : \qsta$}
\end{prooftree}
By induction, $\Gamma^* \vdash A^* : \reltype \qsta^* A A'$.

By conversion, this gives $\Gamma^* \vdash A^* : \eqtype A A'$.

We also have $(\Gamma, x{:} TA)^* \vdash B^* : \reltype \qsta^* B B'$.

That gives $\Gamma^*, x{:}TA, x':TA', x^*:\reltype A^* x x' \vdash B^* : \eqtype B
B'$.

Using the abstraction rule, we derive
\begin{align*}
 \Gamma^* &\vdash \lambda x{:}TA\; \lambda x'{:}TA' \; \lambda x^* :
\reltype A^* x x'. B^* \\
\qquad : \Pi x{:}TA \; \Pi x'{:}TA\ \; \Pi x^* : \reltype A^* x
x'. \eqtype B B'
\end{align*}
which can be rewritten as
\[\Gamma^* \vdash (\lambda x : TA. B)^* :
\Pi x{:}TA \; \Pi x'{:}TA\ \; \Pi x^* : \reltype A^* x x'. \; \eqtype
B B'\]
Using the $\eqfun$-constructor, we may derive
\[ \Gamma^* \vdash \eqfun A^* (\lambda x: TA. B)^* : \eqtype
(\qFun A B) (\qFun A' B') \]
The subject of the above judgment is equal to 
\[ (\qFun A (\lambda x:TA.B))^* \]
while the type is convertible to $\reltype \eqsta (\qFun A B) (\qFun A' B')$.
Putting these together using the conversion rule yields
\[ \Gamma^* \vdash (\qFun A (\lambda x:TA.B))^* :
\reltype \qsta^* (\qFun A B) (\qFun A B)'\]
being of the required form.

By replacing $\Pi$ with $\Sigma$, $\qFun$ with $\qSum$, and $\eqfun$
with
$\eqsum$, we may derive from the same hypotheses that
\[ \Gamma^* \vdash (\qSum A (\lambda x:TA.B))^* :
\reltype \qsta^* (\qSum A B) (\qSum A B)'\]
\item[Abstraction] 
If we have to do
\begin{prooftree}
  \AXC{$\Gamma \vdash A : T \qsta$}
  \AXC{$\Gamma, x : T A \vdash B : T \qsta$}
  \AXC{$\Gamma, x : T A \vdash b : T B$}
  \TIC{$\Gamma \vdash \lambda x{:} T A.b \; : \; T (\qFun A (\lambda
    x{:}TA. B))$}
\end{prooftree}
the induction hypotheses yield, with conversion, that
\begin{align*}
  \Gamma^* &\vdash A^* : \eqtype A A'\\
  \Gamma^*, x:TA, x':TA', x^* : \reltype A^* x x'
&\vdash B^* : \eqtype B B'\\
  \Gamma^*, x:TA, x':TA', x^* : \reltype A^* x x'
&\vdash b^* : \reltype B^* b b'\\
\end{align*}

Since $A, A'$ are $U$-types in $\Gamma^*$, and
$\reltype A^* x x' : *$, we can apply the abstraction rule 
three times in a row to see that the context
\[ (\Gamma, x : T A)^* \; = \; \Gamma^*, x: TA, x' :TA', x^*: \reltype
A^* x x' \]
yields typing judgment
\[\vdash \lambda x{:}TA\; \lambda x'{:}TA'\;
\lambda x^* : \reltype A^* x x'. b^*
\;:\; \Pi x{:}TA\; \Pi x'{:}TA'\;
\Pi x^* : \reltype A^* x x'. \reltype B^* b b'\]

The subject of this judgment is equal to $(\lambda x : T A. b)^*$.

The type predicate may be converted as
\begin{align*}
\Pi x{:}TA\; \Pi x'{:}TA'\;
&\Pi x^* : \reltype A^* x x'.\quad \reltype B^* b b'\\
= \Pi x{:}TA\; \Pi x'{:}TA'\;
&\Pi x^* : \reltype A^* x x'. \quad \reltype B^* ((\lambda x{:}TA.b)x)
((\lambda x'{:}TA'.b')x')\\
= \Pi x{:}TA\; \Pi x'{:}TA'\;
&\Pi x^* : \reltype A^* x x'. \\
&\reltype ((\lambda x:TA. B)^* x x' x^*) ((\lambda x{:}TA.b)x)
((\lambda x'{:}TA'.b')x')\\
&\hspace{-2cm}= \reltype (\eqfun A^* (\lambda x:TA. B)^*)
(\lambda x{:}TA. b) (\lambda x{:}TA. b)'\\
&\hspace{-2cm}= \reltype (\qFun A (\lambda x:TA.B))^*
(\lambda x{:}TA. b) (\lambda x{:}TA. b)'
\end{align*}
which is of the form \eqref{e:ext}, as desired.
\item[Application] 
If the derivation ends with
\begin{prooftree}
  \AXC{$\Gamma\phantom{, x{:}TA } \vdash A : T \qsta$}
  \noLine
  \UIC{$\Gamma, x{:}TA \vdash B:T\qsta$}
  \AXC{$\Gamma \vdash f : T(\qFun A (\lambda x{:}TA. B))$}
  \AXC{$\Gamma \vdash a : T A$}
  \TIC{$\Gamma \vdash f a : B [a/x]$}
\end{prooftree}
We thus have that
that $A$ ($A'$) and $B$ ($B'$) are $U$-types in
$\Gamma$ ($\Gamma'$) and $\Gamma, x{:}TA$ ($\Gamma', x':TA'$),
respectively.

The induction hypotheses give us
\begin{align*}
\Gamma^* &\vdash A^* : \eqtype A A'\\
(\Gamma, x:TA)^* &\vdash B^* : \eqtype B B'\\
\Gamma^* &\vdash f^* : \reltype (\qFun A (\lambda x{:}T A. B))^* f f'\\
\Gamma^* &\vdash a^* : \reltype A^* a a'
\end{align*}
We may rewrite the type of $f^*$ as
\begin{align*}
&\reltype (\qFun A (\lambda x{:}T A. B))^* f f' \\
&= \Pi x{:}TA\; \Pi x'{:}TA'\; \Pi x^* : \reltype A^* x x'.\\
&\quad\qquad \reltype ((\lambda x{:}TA.B)^* x x' x^*) (f x) (f' x')\\
&= \Pi x{:}TA\; \Pi x'{:}TA'\; \Pi x^* : \reltype A^* x x'.\\
&\quad\qquad \reltype ((\lambda x{:}TA\;\lambda x'{:}TA'\;
\lambda x^*:\reltype A^* x x'.\; \reltype B^*) x x' x^*) (f x) (f'
x')\\
&= \Pi x{:}TA\; \Pi x'{:}TA'\; \Pi x^* : \reltype A^* x x'.
\reltype B^* (f x) (f' x')
\end{align*}
Working in $\Gamma^*$, we now apply $f^*$ to $a, a',a^*$
(which types are $T A, T A', \reltype A^* a a'$,
respectively), in order to obtain
\[ f^* a a' a^* : \reltype B^* (f x) (f' x') [a/x,a'/x',a^*/x^*],\]
where we have used the hypotheses on $A^*$ and $B^*$
in validating application typing rule.

Since the sets of primed, starred, and vanilla variables are disjoint,
and every variable in $f'$ is primed, while every veriable in $f$
vanilla,
we may rewrite the above as
\[ f^* a a' a^* : \reltype B^* [a/x, a'/x',a^*/x^*] (f a) (f' a') \]
By the substitution lemma,
\[ B^* [a/x, a'/x', a^*/x^*] = (B[a/x])^*\] 
We may thus rewrite the above judgment as
\[ \Gamma^* \vdash (f a)^* : \reltype B[a/x]^* (f a) (f a)' \]
as required.
\item[Pairing] Given a derivation
\begin{prooftree}
  \AXC{$\Gamma\phantom{, x{:}TA } \vdash A : T \qsta$}
  \noLine
  \UIC{$\Gamma, x{:}TA \vdash B:T\qsta$}
  \AXC{$\Gamma \vdash a : T A$}
  \AXC{$\Gamma \vdash b : T B [a/x]$}
  \TIC{$\Gamma \vdash (a,b) : T (\qSum A (\lambda x:TA. B))$}
\end{prooftree}
we have 
\begin{align*}
  \Gamma^* &\vdash a^* : \reltype A^* a a'\\
  \Gamma^* &\vdash b^* : \reltype B[a/x]^* b b'
\end{align*}
We also have
\begin{align*}
  &\reltype (\qSum A (\lambda x{:} TA. B))^* (a,b) (a',b') \\
  &= \reltype (\eqsum A^* (\lambda x : TA. B)^*) (a,b) (a',b')\\
  &= \Sigma a^* : \reltype A^* \pi_1 (a,b) \pi_1 (a',b').\\
   &\qquad\reltype ((\lambda x{:} TA. B)^* \pi_1 (a,b) \pi_1 (a',b') a^*)
  \pi_2 (a,b) \pi_2 (a',b')\\
  &= \Sigma a^* : \reltype A^* a a'.
  \reltype ((\lambda x{:} TA.B)^* a a' a^*) b b'\\
  &= \Sigma a^* : \reltype A^* a a'.
  \reltype (B^* [a,a',a^*/x,x',x^*]) b b'\\
  &= \Sigma a^* : \reltype A^* a a'. 
  \reltype B [a/x]^* b b'
\end{align*}
Using the pairing rule, we see that $(a^*,b^*)$ can be given the type
derived above.  So by conversion, we find
\[ \Gamma^* \vdash (a,b)^* : \reltype (\qSum A (\lambda x{:} TA. B))^*
(a,b) (a,b)' \]
as required.
\item[Projections] 
Given 
\begin{prooftree}
  \AXC{$\Gamma\phantom{, x{:}TA } \vdash A : T \qsta$}
  \noLine
  \UIC{$\Gamma, x{:}TA \vdash B:T\qsta$}
  \AXC{$\Gamma \vdash p : T (\qSum A (\lambda x{:}TA. B))$}
  \BIC{$\Gamma \vdash \pi_1 p : T A$}
  \noLine
  \UIC{$\Gamma \vdash \pi_2 p : T B[\pi_1 p/x]$}
\end{prooftree}
we get, by induction hypothesis, that
\[\Gamma^* \vdash p^* : \reltype (\qSum A (\lambda x{:} TA. B))^* p
p'\]
By the same computation as the previous case, we see that that the
type of $p^*$
above is convertible to
\[\Sigma a^* : \reltype A^* (\pi_1 p)  (\pi_1 p').
\reltype B^*[\pi_1 p, \pi_1 p', a^* / x, x', x^*] (\pi_2 p) (\pi_2
p') \]
But then we have
\[ \pi_1 p^* : \reltype A^* \pi_1 p \pi_1 p' \]
\[ \pi_2 p^*: \reltype B^*[\pi_1 p, \pi_1 p', \pi_1 p^*/x,x',x^*] (\pi_2 p)
(\pi_2 p') \]
The first judgment above already has the form required.  As for the
second, we use the substitution lemma to rewrite it as
\[ (\pi_2 p)^* : \reltype B [\pi_1 p/x]^* (\pi_2 p) (\pi_2 p)' \]
and this too obeys the form of \eqref{e:ext}.
\item[Conversion] Suppose
  \begin{prooftree}
    \AXC{$\Gamma \vdash M : A$}
    \AXC{$\Gamma \vdash B : T \qsta$}
    \AXC{$A = B$}
    \TIC{$\Gamma \vdash M : B$}
  \end{prooftree}
By induction hypothesis, we have
\[ \Gamma \vdash M^* : \reltype A^* M M' \]

The fact that $A = B$, entails, for deep typographical reasons, that
$A' = B'$.

Hence by conversion, we have that $M : B$ as well as $M' : B'$.

But we also have that $B : U$, so that $B^* : \reltype \qsta^* B B'$,
or equivalently $B^* : \eqtype B B'$.

These facts yield that $\reltype B^* M M' : *$.

By Proposition \ref{c:conv}, $\reltype A^* M M' = \reltype B^* M M'$.

$\Gamma^* \vdash M^* : \reltype B^* M M' \qedhere$
\end{description}
\end{proof}

\section{Higher dimensions}

The theorem in the previous section can be extended to higher
dimensions.
For example, for the groupoid level, we can write an IIRD defining
terms
\begin{align*} &\eqtype_2 : \Pi A B: U. \eqtype A B \to \eqtype A B \to \sta\\
 &\reltype_2 : \Pi A B: U \Pi e  e' : \eqtype A B. \eqtype_2 A B e e'
 \to \sta
\end{align*}
We can follow the same steps as before, and prove extensionality of
the 1-dimensional $\lstaue$ terms as well.

In the next dimension, we would define terms
\begin{align*} &\eqtype_3 : \Pi A B: U\; \Pi e e' : \eqtype A
  B.\; \eqtype_2 e e' \to \eqtype_2 e e' \to \sta\\
 &\reltype_3 : \Pi A B: U\; \Pi e  e' : \eqtype A B\;
 \Pi \eps \eps' : \eqtype_2 A B e e'. \; \eqtype_3 A B e e' \eps \eps'
 \to \sta
\end{align*}

It is clear that this method could be continued indefinitely.
Once we build a countable sequence of such IIRD types, we get
a type theory in which the extensionality of every term is
witnessed by a higher-dimensional instance of that term.

It is easier to look at the pattern of the definitions required to
ascend extensionality from one dimension to the next.  We can then
cover all dimensions at once by making this pattern part of our
universe.

On the next page, we give a double IIRD in which the equivalence and
relation types are fully internalized into the universe.  The notation
\[\qFun x : A. \ul{\qquad}\]
\[\qSum x : A. \ul{\qquad}\]
is used there to denote the terms
\[\qFun A (\lambda x: TA. \ul{\qquad})\]
\[\qSum A (\lambda x: TA. \ul{\qquad})\]

\section{Future work}
In order to reason about extensional identity type \emph{within} the
system itself, it could feel good to internalize the above theorem into the
language of type theory.  This will require reflecting not only the
type level but also the syntax of terms.  Furthermore, one would need
to simultaneously treat symmetry and transitivity, together with the
interaction between them and everything else.

\newcommand{\spaceit}{\hspace{-2cm}|\;}
\newcommand{\spaceitm}{\hspace{-2cm}}
\newcommand{\spaceitj}{\hspace{-1.15cm}}

\begin{align*}
\mathsf{Inductive}\ &U : * \;\; :=\\
&|\ \qsta : U\\
&|\ \qFun : \Pi A:U.\; (TA \to U) \to U\\
&|\ \qSum : \Pi A:U.\; (TA \to U) \to U\\
&|\ \qEq : U \to U \to U\\
&|\ \qRel : \Pi \setof{A B} : U.\; T [\qEq A B] \to T A \to T B \to U \hspace{-4cm}\\
\mathsf{with} \quad &T : U \to \sta \;\; :=\\
&T(\qsta) = U\\
&T(\qFun A B) = \Pi a:TA.\; T[Ba]\\
&T(\qSum A B) = \Sigma a:TA.\; T[Ba]\\
&T(\qEq A B) =  [\mathsf{Inductive} &&\hspace{-1cm}\eqtype : U \to U \to \sta\\
&&&\spaceit\eqsta : \eqtype \qsta \qsta\\
&&&\spaceit\eqfun : \Pi \setof{A} \setof{A'} \; \Pi A^* : \eqtype A A'\\
&&&\spaceitj\Pi \setof{B} \setof{B'} \; \Pi B^* : (\Pi a a'
a^*.\; \eqtype (B a) (B' a')).\\
&&&\hspace{1.5cm}  \eqtype (\qFun A B) (\qFun A' B')\\
&&&\spaceit\eqsum : \Pi \setof{A} \setof{A'} \; \Pi A^* : \eqtype A A'\\
&&&\spaceitj\Pi \setof{B} \setof{B'} \; \Pi B^* : (\Pi a a'
a^*.\; \eqtype (B a) (B' a')).\\
&&&\hspace{1.5cm}  \eqtype (\qSum A B) (\qSum A' B')\\
&&&\spaceit\eqEq : \Pi \setof{A A'} A^* \setof{B B'} B^*.\quad  \eqtype (\qEq A B) (\qEq
A' B')\\
&&&\spaceit\eqRel : \Pi A A' A^* B B' B^* e e' e^* a a' a^* b b' b^*.
\quad \eqtype (\qRel e a b) (\qRel e' a' b')\\
&\hspace{2cm}\mathsf{ with }  && \hspace{-1.5cm}\reltype \setof{A B: U}:
\eqtype A B \to T  A \to T B \to U\\
&
&&\spaceitm\reltype (\eqsta) A B = \qEq A B\\
&&&\spaceitm\reltype (\eqfun A^* B^*) f f' = \qFun x{:}A \qFun\! x'{:}A' \qFun\! x^* :
\reltype A^* x x'. \\
  &&&\hspace{2cm}
  \reltype (B^* x x' x^*) (f x) (f' x')\\
  &&&\spaceitm\reltype (\eqsum A^* B^*) p p' =
  \qSum x^* : \reltype A^* (\pi_1 p) (\pi_1 p'). \\
  &&&\hspace{2cm}
  \reltype (B^* (\pi_1 p) (\pi_1 p') x^*)
  (\pi_2 p) (\pi_2 p')\\
  &&&\spaceitm\reltype (\eqEq A A' A^* B B' B^*) e e' =
  \qFun a a' a^*\; \qFun\! b b' b^*.\;\\
  &&& \hspace{2cm} \qEq\! (\qRel e a b) (\qRel e' a' b')\\
  &&&\spaceitm\reltype (\eqRel A A' A^* B B' B^* e e' e^* a a' a^* b b' b^*)
  \gamma \gamma' =\\
  &&& \hspace{2cm} \qRel\! (e^*a a' a^* b b' b^*) \gamma \gamma'\\
&\hspace{2cm} ] A B\\
&T (\qRel e a b) = T(\reltype e a b)
\end{align*}

\bibliography{ett1.bib}

\end{document}

%% file: amslambda.tex
\usepackage{amsmath}
\usepackage{stmaryrd}
\usepackage{MnSymbol}
\usepackage{bussproofs}
\EnableBpAbbreviations

\DeclareFontFamily{OT1}{pzc}{}
\DeclareFontShape{OT1}{pzc}{m}{it}{<-> s * [1.10] pzcmi7t}{}
\DeclareMathAlphabet{\mathpzc}{OT1}{pzc}{m}{it}
\newcommand{\ul}[1]{\underline{#1}}
\newcommand{\types}{\mathbb{T}}

\newcommand{\terms}{\Lambda}

\newcommand{\then}{\;\Longrightarrow\;}
\newcommand{\comment}[1]{}

\newcommand{\RA}{\Rightarrow}

\newcommand{\sta}{*}
\newcommand{\lstar}{{\lambda\!\!\:\sta}}
\newcommand{\lstars}{\lstar}
\newcommand{\lstarss}{{\lambda\!\sta}}

\newcommand{\qfun}{\mathbin{{\ooalign{$\medcircle$\cr\raise.29ex
\hbox{$\mkern1.9mu\scriptscriptstyle{\to}$}\cr}}}}

\newcommand{\qlam}{\mathsf{Lam}}
\newcommand{\qapp}{\mathsf{App}}
\newcommand{\qpair}{\mathsf{Pair}}
\newcommand{\qproj}{\mathsf{Proj}}

\newcommand{\ee}[1]{\simeq_{#1}}
\newcommand{\eqv}{\ee{}}
\newcommand{\env}{\mathsf{Env}}
\newcommand{\btype}{\mathbf{Type}}
\newcommand{\lsimple}{\lambda_{\to}}
\newcommand{\setof}[1]{\{{#1}\}}

\newcommand{\semof}[1]{\llbracket{#1}\rrbracket}
\newcommand{\sse}{\subseteq}

\newcommand{\refl}[1]{{\mathsf{r}({#1})}}

\newcommand{\ol}[1]{\overline{#1}}

\newcommand{\qFun}{\mathbin{{\ooalign{$\bigcircle$\cr\raise.037ex
\hbox{$\mkern3.1mu\scriptstyle{\Pi}$}\cr}}}}
\newcommand{\qSum}{\mathbin{{\ooalign{$\bigcircle$\cr\raise.036ex
\hbox{$\mkern3.5mu\scriptstyle{\Sigma}$}\cr}}}}

\newcommand{\qsta}{\mathord{\oast}}
\newcommand{\qU}{\mathbin{{\ooalign{$\bigcircle$\cr\raise.037ex
\hbox{$\mkern5mu\scriptstyle{\mathtt{U}}$}\cr}}}}
\newcommand{\qT}{\mathbin{{\ooalign{$\bigcircle$\cr\raise.037ex
\hbox{$\mkern5mu\scriptstyle{\mathtt{T}}$}\cr}}}}
\newcommand{\qEq}{\mathbin{{\ooalign{$\bigcircle$\cr\raise.15ex
\hbox{$\mkern5mu\scriptstyle{\ee{}}$}\cr}}}}
\newcommand{\qRel}{\mathbin{{\ooalign{$\bigcircle$\cr\raise.2ex
\hbox{$\mkern5mu\scriptstyle{\sim}$}\cr}}}}
\newcommand{\lstau}{{\lstar U}}

\newcommand{\eqsta}{\refl{\qsta}}
\newcommand{\eqfun}{{\qFun}\!{}^*}
\newcommand{\eqsum}{{\qSum}\!{}^*}
\newcommand{\eqtype}{\mathpzc{Eq}} 
\newcommand{\reltype}{\mathpzc{Rel}} 
 
\newcommand{\lstaue}{\lstau\!{\ee{}}}
\newcommand{\eps}{\epsilon}
\newcommand{\eqEq}{{{\qEq}\!{}^*}}
\newcommand{\eqRel}{{{\qRel}\!{}^*}}

\comment{

}

\renewcommand{\refl}[1]{\mathsf{r}({#1})}